\documentclass[journal,twocolumn]{IEEEtran}
\usepackage[left=1.45cm, right=1.45cm, top=1.45cm, bottom = 1.45cm]{geometry}
\usepackage{stmaryrd}
\usepackage{cite}
\usepackage{float}
\usepackage{amsmath,amssymb,amsfonts}
\usepackage{algorithmic}
\usepackage{graphicx} 
\usepackage[table]{xcolor}
\usepackage{textcomp}
\usepackage{lipsum}
\usepackage{xcolor}
\usepackage{booktabs}
\def\BibTeX{{\rm B\kern-.05em{\sc i\kern-.025em b}\kern-.08em
    T\kern-.1667em\lower.7ex\hbox{E}\kern-.125emX}}
\usepackage{subfigure}
\usepackage{mathtools}
\usepackage{multirow}
\usepackage{lettrine}
\usepackage[ruled,vlined,linesnumbered]{algorithm2e}
\usepackage{babel,blindtext}
\usepackage[spaces,hyphens]{url}

\usepackage{tabularx}
\usepackage[justification=centering]{caption}
\usepackage{amsthm}

\usepackage{nccmath}
\usepackage{booktabs,tikz}

\usepackage{lipsum, babel}
\usepackage{soul}

\begin{document}

\title{\huge{Optimizing Information Freshness in  RIS-assisted
NOMA-based IoT Networks}}

\author{
 Ali~Muhammad, Mohamed Elhattab, Mohamed Amine Arfaoui,  and~Chadi~Assi
 
}
\maketitle
\IEEEpeerreviewmaketitle

\begin{abstract}
This paper investigates the benefits of integrating reconfigurable intelligent surface (RIS) on minimizing the average sum age of information (AoI) in uplink non-orthogonal multiple access-based Internet-of-Things (IoT) networks. In this setup, an optimization problem is formulated to optimize the RIS configuration, the transmit power per IoT device and the clustering policy of IoT devices. The formulated problem is a mixed-integer non-convex one, and in order to solve it we obtain first the RIS configuration by adopting a semi-definite relaxation (SDR) approach. Afterwards, the joint power allocation and user-clustering problem is solved using the concept of bi-level optimization and is decomposed into an outer user clustering problem and an inner power allocation problem. Optimal closed-form expressions are derived for the inner problem and the Hungarian method is employed to solve the outer one. Numerical results demonstrate the performance superiority of our approach.
\end{abstract}
\vspace{-0.2cm}
\section{Introduction}
 Age of Information (AoI) offers a rigorous way to quantify the information freshness for Internet-of-Things (IoT) applications, which rely on the timely delivery of the status information of physical processes, e.g., environment monitoring in sensor networks \cite{9215013}. Out-dated information updates are undesirable as being inconsistent with the current state of the system and may provoke unreliable/erroneous decisions. AoI can be defined as the elapsed time since the most recent delivered status message was generated \cite{kaul2012real}. Most literature on AoI analysis mainly focused on the orthogonal multiple access (OMA) schemes, i.e., a single user could transmit its information update on a particular resource block (time and/or frequency) \cite{maatouk2019minimizing}. However, with the quest of  massive connectivity and high spectral efficiency requirements of IoT applications, non-orthogonal multiple access (NOMA) appears to be a promising scheme and is a key access for future wireless networks, accounting on its superior spectral efficiency performance against its contemporary OMA techniques~\cite{liu2020uplink}.  
 \par Recent investigations aiming to unleash the full potentials of NOMA towards optimizing the AoI in real-time applications unveiled that the timely delivery of information update messages can be severely compromised by the impact of the wireless channels \cite{ 9477193,  wang2021optimizing}. In a realistic sense, the highly random and uncontrollable behaviour of wireless communication environments impede the timely delivery of information update messages. Typically, a strong communication link between a source and a destination is unattainable due to the channel impairments and blockages.
\par Owing to reconfigurable intelligent surface (RIS), a disruptive technology that is expected to enable next-generation wireless networks \cite{9122596}, it has become possible to construct a strong channel between the source and destination by alleviating the propagation impairments of the wireless environments. Specifically, RIS consists of a planar array of passive elements, each of which has the ability to independently tune the phase-shift of the impinging waves. Therefore, the signals transmitted within the wireless environment can be controlled, and through a proper adjustment of the phase-shifts of all the RIS elements, the desired signals at the points of interest can be enhanced \cite{Marco_JSAC}.
\par There have been some efforts recently to explore the benefits of RIS in NOMA-based systems to improve network reliability \cite{Habder_2021_Latency}, coverage range  \cite{9462949}, network sum-rate \cite{Elhattab_JSAC} and secrecy performance \cite{9385957}. However, the proposed approaches may not necessarily be optimal in the context of information freshness. Recently, the works \cite{samir2020optimizing,muhammad2021age} have exploited the use of RIS for the AoI optimization. The authors of \cite{samir2020optimizing} explored the benefits brought by aerial-RIS on AoI. However, the authors assumed that a single IoT device is scheduled in each time-slot and the impact of the direct channels from the BS to the IoT devices (IoTDs) were neglected. The analysis of AoI in \cite{muhammad2021age} involved the direct channels from the BS to the IoTDs and considered optimizing the RIS configuration for a finite number of IoTDs. Even though both of these works highlighted the significance of leveraging RIS to minimize the AoI, the scope is limited to OMA schemes. 
\par Motivated by the above, we aim in this paper to investigate the envisioned benefits of integrating RIS in NOMA-based systems, where multiple IoTDs are transmitting time-sensitive information updates to the BS. An RIS is deployed to enhance the quality of the wireless channels and status updates are considered to be delivered successfully if each of the links in the network is not in outage. For this setup, a joint RIS phase-shift matrix, user-clustering policy, and user transmit power is formulated as an optimization problem with the objective of minimizing the average sum AoI. The formulated problem is a mixed integer non-convex problem, which is difficult to solve. In order to tackle this issue, we first obtain the RIS phase-shift matrix that maximizes the minimum channel gains of all weak NOMA IoTDs. Afterwards, the original problem is reformulated as a bi-level optimization problem, comprising an outer user clustering problem and an inner power allocation problem. The inner problem is a feasibility condition problem in which we derive the feasible range for the user transmit power. Meanwhile, the outer problem is a classical linear assignment problem, which is solved through the Hungarian method. 
\vspace{-0.4cm}
\section{System Model}
\vspace{0.1cm}
\label{sysmodel}
\subsection{Network Model}

\begin{figure}[t]
\centering
\includegraphics[width=0.6\linewidth]{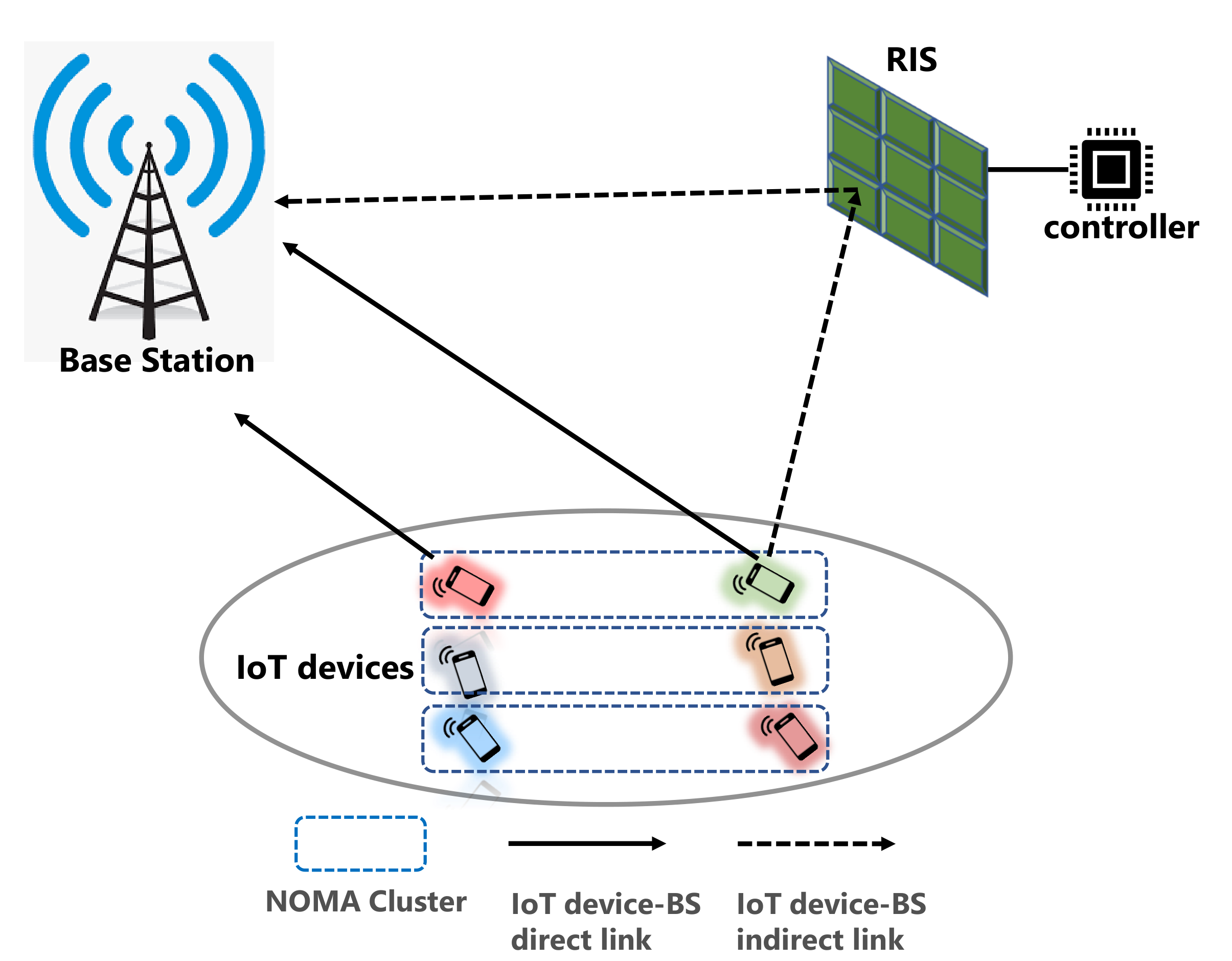}
\caption{An illustration of our system model.}
\label{systemmodel}
\end{figure} 
We consider the uplink IoT network depicted in Fig. \ref{systemmodel}, which consists of one BS and $2I$ IoTDs that provide time-stamped status update information to the BS. The time is divided into time-slots with a slot index $t \in \mathcal{T}$, where $\mathcal{T} = \{1, 2, \dots , T\}$, such that $T$ denotes the time horizon of this discrete-time system. Moreover, let $\mathcal{I} = \{1, 2, \dots ,  2I\}$ denote the set of IoTDs. The IoTDs are divided into two disjoint sets with equal size $I$ based on their channel gains \cite{9140040}. The first set, denoted by $\mathcal{S}$, contains the $I$ IoTDs with the highest channel gains, which are referred to as strong IoTDs. On the other hand, the second set, denoted by $\mathcal{W}$, contains the $I$ IoTDs with the lowest channel gains, which are referred to as weak IoTDs. The $2I$ IoTDs are grouped into $I$ disjoint clusters by pairing exactly one strong IoTD from $\mathcal{S}$ with exactly one IoTD from $\mathcal{W}$. Each cluster communicates with the BS using the uplink NOMA scheme. Moreover, in order to eliminate the inter-cluster interference experienced at the BS, the different clusters communicate with the BS simultaneously over orthogonal frequency sub-carriers \cite{9140040}. Due to impurities and the obstacles of the wireless propagation environment, the existence of a strong direct line-of-sight (LoS) communication link between each weak IoT device and the BS is difficult to obtain. For this purpose, an RIS equipped with $L$ reflecting elements is assumed to be deployed in order to assist the uplink transmission from the weak NOMA IoTDs to the BS by passively relaying the status update information to the BS. The BS continuously controls the phase-shift of the reflecting elements in order to maintain the required quality of service of the weak IoTDs.  Let $ \boldsymbol \Phi(t)=  {\rm diag} \left( \exp \left[j \boldsymbol{\theta}(t)\right] \right) \in \mathbb{C}^{L \times L}$ be the RIS's diagonal phase-shift matrix in the $t$th time-slot, where $\boldsymbol{\theta}(t) = \left[\theta_1(t), \dots, \theta_L(t)\right]$ and, for all $l \in L$, $\theta_l (t) \in [0,2\pi)$ is the phase-shift of the $l$th reflecting element of the RIS. 
\vspace{-0.4cm}
\subsection{Channel Model and SINR Analysis}
To properly illustrate the signal model at BS, a single NOMA cluster is considered. Consider a cluster of IoTDs $(s,w)$ $\in  \mathcal{S}\times\mathcal{W}$, where $s$ and $w$ represent the strong and weak IoTDs in the considered NOMA, respectively. The channel gain between the RIS and the BS is denoted by $\boldsymbol h_{R \rightarrow b}(t) \in \mathbb{C}^{L \times 1}$, between the weak user $w$ and the RIS is denoted by ${\boldsymbol h_{w \rightarrow  R}}(t) \in \mathbb{C}^{L \times 1}$, and between user $x$ and the BS, for $x \in \{s,w\}$, by $h_{x \rightarrow b}(t)  \in \mathbb{C}$. All channel gains consist of both the small-scale and the large-scale fading, which are given as $\boldsymbol h_{R \rightarrow  b}(t) =   \hat {\boldsymbol h}_{ R \rightarrow  b}(t) \Delta_{R \rightarrow b}$, $\boldsymbol  h_{ w \rightarrow  R}(t) =  \hat {\boldsymbol h}_{ w \rightarrow  R}(t) \Delta_{w \rightarrow R}$, and $h_{x \rightarrow b}(t) = \hat h_{x \rightarrow b}(t) \Delta_{x \rightarrow b}$, for $x \in \{s,w\}$, where $\hat { \boldsymbol h}_{ R \rightarrow  b}(t)$, $\hat { \boldsymbol h}_{ w \rightarrow  R}(t)$, and $\hat {h}_{x \rightarrow  b}(t)$ represent the small-scale fading coefficients between the RIS and the BS, between the weak IoTD $w$ and the RIS, and between the IoTD $x \in \{s,w\}$ and the BS respectively. The large-scale fading coefficients $\Delta_{R \rightarrow b}$, $\Delta_{w \rightarrow R}$, and $\Delta_{x \rightarrow b}$, for $x \in \{s,w\}$, are modeled as
$\Delta_{R \rightarrow b} = \sqrt{\gamma_0 d_{R \rightarrow b} ^{- \eta_{R \rightarrow b}}}$, $\Delta_{w \rightarrow R} = \sqrt{\gamma_0 d_{w \rightarrow R} ^{- \eta_{w \rightarrow R}}}$ and
$\Delta_{x \rightarrow b} = \sqrt{\gamma_0 d_{x \rightarrow b} ^{- \eta_{x \rightarrow b}}}$, where $\gamma_0$ is the path-loss average channel power gain at a reference distance $d_0 =$1m, $\eta_{k}$, for $k \in \left\{R \rightarrow b,w \rightarrow R,x \rightarrow b \right\}$ is the path-loss exponent of the wireless link $k$, while $d_{k}$ represents the distance for the communication link $k$. The small scale fading of the direct links between the IoTDs and the BS is modeled as a Rayleigh fading with a zero mean and a unit variance \cite{elhattab2101reconfigurable}. Meanwhile, the communication links between the RIS and the BS and between the weak IoTD $w$ and the RIS are considered to have LoS components. These links experience a small-scale fading that is modeled as a Rician fading. Hence, the small-scale fading between the BS and RIS is given as \cite{elhattab2101reconfigurable}
\begin{equation}
    \hat {\boldsymbol h}_{R \rightarrow b}(t) = \sqrt{\frac{K_1}{K_1 + 1}}  \tilde {\boldsymbol h}_{R \rightarrow b}(t) + \sqrt{\frac{1}{K_1 + 1}}  \bar {\boldsymbol h}_{R \rightarrow b}(t),
    \label{ssf1}
\end{equation}

\noindent
{where $K_1$ is the Rician factor, and $\tilde {\boldsymbol h}_{R \rightarrow b}(t)$ and $\bar {\boldsymbol h}_{R \rightarrow b}(t)$  are the deterministic LoS and Rayleigh fading components. Note that, the $\hat {\boldsymbol h}_{ w \rightarrow  R}(t)$ can be obtained similarly.} 
\par Let $p_s(t)$ and $p_w(t)$ denote the transmit power of the IoTD $s$ and $w$ in cluster $(s,w)$ at time-slot $t$, respectively. In uplink NOMA, the BS receives the signals from the IoTDs $s$ and $w$, respectively. Then, it sequentially applies successive interference cancellation (SIC) to decode the signals of both IoTDs, where the decoding order is determined based on the channel gains of the IoTDs. In fact, the BS starts by decoding the signal of the strong IoTD $s$ while considering the signal of the weak IoTD $w$ as an interference. Consequently, the received signal to noise plus interference ratio (SINR) of IoT user $s$ at the BS in time-slot $t \in T$ can be expressed as
\begin{equation} 
  \gamma_{s}(t) =
\frac{p_s(t) |h_{s \rightarrow b}(t) |^2} {p_w(t)| \boldsymbol h_{w \rightarrow R}(t) \boldsymbol \Phi(t) \boldsymbol h^H_{R \rightarrow b}(t) + h_{w \rightarrow b}(t) |^2+\sigma^2},
 \label{snrconstraint}
\end{equation}

\noindent
where $\sigma^2$ is the variance of zero-mean additive Gaussian noise. Afterwards, the decoded signal of the strong IoTD $s$ is removed from the received signal at the BS and, then, the signal of IoTD $w$ will be decoded without any interference. Hence, the SINR of the IoTD $w$ at the BS in time-slot $t \in T$ can be expressed as
\begin{equation} 
  \gamma_{w}(t) =
\frac{p_w(t)| \boldsymbol h_{w \rightarrow R}(t) \boldsymbol \Phi(t) \boldsymbol h^H_{R \rightarrow b}(t) + h_{w \rightarrow b}(t) |^2}{\sigma^2}.
 \label{snrconstraint2}
\end{equation}
\vspace{-0.5cm}
\subsection{AoI Modeling}
The AoI illustrates how old is the information of a source from the destination's perspective and characterizes the inter-delivery time of arrivals and the latency. For all $i \in \mathcal{I}$ and $t \in \mathcal{T}$, we denote the AoI of the $i$th IoTD at the BS in $t$th  time-slot as $A_{i}(t)$. A successful delivery of a packet at the BS in a given time slot $t$ is conditioned on the case that the SINR of the channel between the IoTD and the BS is above a given threshold denoted as $\gamma_{\rm{th}}$. Precisely, if a packet of the $i$th IoTD is  successfully delivered at the BS, the corresponding AoI in the same time-slot is given by  $A_{i}(t)= 1$.  Conversely, if the transmission remains unsuccessful, the AoI value will increase by $1$, i.e., $A_{i}(t)=A_{i}({t-1})+1$.  In this work, and similar to \cite{wang2021optimizing}, a \textit{generate at will} model is considered, where a packet is generated at the start of a time-slot and the transmission occurs in the same time-slot.  The evolution of AoI of the $i$th IoTD is given by
\begin{equation}
 A_{i}({t}) = 
 \begin{cases}
  {1}, & \text{if  }    U(I-i)  \gamma_{s}(t) \text{+} \\
  & \,  U(i- I)  \gamma_{w}(t)  
  \geq \gamma_{\mathrm{th}}, 
    \\
  A_{i}({t-1})+1, &\text{otherwise,}
 \end{cases}
 \label{mainage}
\end{equation}
$U(x)$ is a unit step function (= 1 if $x > 0$, 0 otherwise).
\vspace{-0.1cm}
\section{Problem Formulation and Solution Approach}
\label{formulationprob}
\vspace{-0.1cm}
To ensure the received information at the BS is fresh, we aim to minimize the average sum AoI of all IoTDs. Over the entire horizon $T$, we optimize the RIS configuration, the user-clustering, and the user transmit power. This framework is formulated as an optimization problem, which is written as:
\allowdisplaybreaks
\begin{subequations} 
\begin{align}
\text{$\mathcal {OP} $: } & \min_{ \substack{\boldsymbol B,
\boldsymbol \Phi,\\ \boldsymbol{p_w}, \boldsymbol{p_s}}} \dfrac{1}{T} \sum_{t<T} \Big[\dfrac{1}{2I} \sum_{s \in \mathcal{S}} \sum_{w \in \mathcal{W}} \Big\{  A_{s}(t) + A_{w}(t) \Big\} b_{s,w}(t) \Big] \\
\text{s.t.  \,\,\, 
} & \label{constheta} \text{$\theta_l(t) \in [0,2\pi), \,\,\, \quad \,\, \,\,\,\,\forall \,\, l \in \mathcal{L},t \in T$, } 
\\ 
&\label{consb1}  b_{s,w}(t)  \in \{0,1\}, \,\,\,\,\, \quad \forall w \in \mathcal{W}, s \in \mathcal{S}, t \in \mathcal{T}, \\
&\label{consbnf1} \sum_{s=1}^ {I} b_{s,w}(t) \leq 1, \,\,\,\,\,\quad \forall w \in \mathcal{W},  t \in \mathcal{T}, \\
&\label{consbnf2} \sum_{w=1}^ {I} b_{s,w}(t) \leq 1, \,\,\,\,\, \quad \forall s \in \mathcal{S},  t \in \mathcal{T}, \\ 
&\label{consbnf3} p_w(t), p_s(t) \leq P_{\max}, \,\forall s \in \mathcal{S}, w \in \mathcal{W}, t \in T,
\end{align}
\end{subequations}
where $b_{s,w}(t) \in \{0,1\}$  is the clustering decision variable, i.e., $b_{s,w}(t)=1$ indicates that IoTD $s \in \mathcal{S}$ is paired with IoTD $w \in \mathcal{W}$ at time-slot $t$, and $b_{s,w}(t)=0$, otherwise, $P_{\max}$ is the IoTD power budget, $\boldsymbol B = \{ b_{s,w} | (s,w)  \in \mathcal{S}\mathcal{W}\}$, $\boldsymbol{p_w} = \{p_w | w \in \mathcal{W}\}$, and $\boldsymbol{p_s} = \{p_s | s \in \mathcal{S}\}$. Constraints (\ref{consbnf1}) and (\ref{consbnf2}) ensure that each strong user $s \in \mathcal{S}$ is paired with at most one weak user $w \in \mathcal{W}$ and and vice versa. Finally, \eqref{consbnf3} ensures that the user transmit power does not exceed its power budget. It can be easily seen that the formulated optimization problem $\mathcal{OP}$ is an  mixed-integer nonlinear programming problem due to the coexistence of the binary clustering matrix $\boldsymbol{B}$ and the continuous variables ($\boldsymbol{\theta}, \boldsymbol{p}_w, \boldsymbol{p}_s$), and therefore, it is difficult to solve directly. 
\vspace{-0.1cm}
\subsection{Solution Approach}
One can observe from problem $\mathcal{OP}$ that it is difficult to obtain the optimal power control, the optimal user clustering and the optimal RIS phase-shift matrix jointly. Consequently, we first obtain the RIS phase-shift matrix that maximizes the minimum channel gains of the weak IoTDs. Afterwards, the obtained RIS phase-shift matrix is injected into the original problem $\mathcal{OP}$ and the resulting problem becomes then a joint user clustering and power control problem which is solved using the concept of bi-level optimization. 
\subsubsection{RIS Phase-Shift Matrix} The RIS phase-shift matrix optimization problem is given as
\begin{subequations} 
\begin{align}
\text{$\mathcal {OP} $1: } & \max_{ 
\boldsymbol \Phi} \min_{ 
 w  \in \mathcal{W}}    {| \boldsymbol h_{w \rightarrow R}(t) \boldsymbol \Phi(t) \boldsymbol h^H_{R \rightarrow b}(t) + h_{w \rightarrow b}(t) |^2}     \\
& \text{s.t.}  \quad \theta_l(t) \in [0,2\pi), \,\,\, \quad \,\, \,\,\,\,\forall \,\, l \in \mathcal{L},t \in T. \label{thetaf}
\end{align}
\end{subequations}
\noindent
Similar to the approach given in \cite{almekhlafi2021joint}, we introduce an auxiliary variable $\zeta$ to solve phase-shift optimization problem. The resulting problem can be then re-written as
\begin{subequations} 
\begin{align}
&\hspace{-0.4cm}\mathcal{OP}2: \max_{ 
 \zeta, \boldsymbol \Phi(t)} \,\,\,  \zeta   \\
& \hspace{-0.4cm}\text{s.t.}\,\,\label{constheta} \zeta \geq0,\\ 
& \hspace{-0.4cm}\label{constheta} | \boldsymbol h_{w \rightarrow R}(t) \boldsymbol \Phi(t) \boldsymbol h^H_{R \rightarrow b}(t) + h_{w \rightarrow b}(t) |^2 \geq \zeta, \, \forall w \in \mathcal{W}, t\in \mathcal{T},\\
&\hspace{-0.4cm}\theta_l(t) \in [0,2\pi), \,\,\, \quad \,\, \,\,\,\,\hspace{3cm}\forall \,\, l \in \mathcal{L},t \in \mathcal{T}.
\end{align}
\end{subequations}
Now, we reformulate the above problem into a rank-one constrained  optimization problem via change of variables and matrix lifting.  Let $\boldsymbol v  \triangleq [ v_1,v_2,....,v_L]^H$, where $v_l=e^{j\theta_l}$ for all $l \in \mathcal{L}$. Thus, for all $l \in \mathcal{L}$, the constraint $\theta_l(t) \in [0,2\pi)$ is equivalent to the unit-modulus constraints, i.e., $|v_l|^2=$1. By applying the change of variables $\boldsymbol h_{w \rightarrow R}(t) \boldsymbol \Phi(t) \boldsymbol h^H_{R \rightarrow b}(t)  = \boldsymbol v^H  \boldsymbol{{Q}}(t)$, where 
$\boldsymbol{{Q}}(t) =\mathrm{diag}(\boldsymbol h^H_{R \rightarrow b}(t))\boldsymbol h_{w \rightarrow R}(t)$, we obtain $|\boldsymbol h_{w \rightarrow R}(t) \boldsymbol \Phi(t) \boldsymbol h^H_{R \rightarrow b}(t) + h_{w \rightarrow b}(t)|^2 = | \boldsymbol v^H \boldsymbol{Q}(t) + h_{w \rightarrow b}(t)|^2 $ = $\bar {\boldsymbol v}^H  \boldsymbol{\Theta} \bar {\boldsymbol v} + |h_{w \rightarrow b}(t)|^2 = {\rm{tr}}(\boldsymbol{\Theta} \bar {\boldsymbol v} \bar {\boldsymbol v}^H) + |h_{w \rightarrow b}(t)|^2$, where  
\begin{align}
    \boldsymbol{\Theta} = \begin{bmatrix} \boldsymbol{Q}(t)\boldsymbol{Q}^H(t) & \boldsymbol{Q}(t)h_{w \rightarrow b}(t) \\ h_{w \rightarrow b}(t)\boldsymbol{Q}^H(t) & 0 \\ \end{bmatrix}, && \bar {\boldsymbol v}= \begin{bmatrix} \boldsymbol v  \\ 1 \end{bmatrix}.
\end{align}
\noindent Now, let  $\boldsymbol{V} \triangleq {\boldsymbol v} \bar {\boldsymbol v}^H$, which needs to satisfy rank$(\boldsymbol V)$ = 1 and $\boldsymbol V \geq$ 0. This rank one constraint is non-convex \cite{elhattab2101reconfigurable}. By dropping this constraint, problem $\mathcal{OP}2$ can be rewritten as
\begin{subequations} \label{eq:subeqf nscdglobalEQ2}
\begin{align}
\label{objop2}
\text{$\mathcal {OP} $3: } & \max_{ \boldsymbol{V},
 \zeta } \,\,\,  \zeta   \\
&\text{s.t.}  \,\,\, tr(\boldsymbol{\Theta}\boldsymbol{V}) + |h_{w \rightarrow b}(t)|^2 \geq  \zeta, \\
& \text{$\boldsymbol{V} \geq$ 0},
  \\
 &\label{vff} [\boldsymbol{V}]_{L,L} = 1.\\
 & \label{constheta} \text{$\zeta \geq$0.} \end{align}
\end{subequations} 
 \par After applying the proposed transformations, the resulted optimization problem turns out to be a convex optimization problem that can be solved by any convex optimization solver such as CVX \cite{elhattab2101reconfigurable,almekhlafi2021joint}. Note that, if the obtained phase-shift matrix does not satisfy the rank-one constraint, the Gaussian randomization (GR) method can be applied to construct a rank-one solution \cite{elhattab2101reconfigurable}. 

\subsubsection{Bi-Level Optimization} We assume that phase-shift matrix is already obtained and served as an input to problem $\mathcal{OP}$.
It can be observed from $\mathcal{OP}$ that the AoI function is independent from the pairing variable $\boldsymbol B$ and it is only a function of the power control policy (see eq. (\ref{mainage})). Precisely, let the set of optimal user clustering and power control scheme, i.e., the solutions of the problem  $\mathcal{OP}$, be denoted by $\left\{\left(b^*_{s,w}(t),  p^*_s(t), p^*_w(t)\right),\,\, s,w \in \mathcal{S} \times \mathcal{W} \right\}$. For all $s,w \in \mathcal{S} \times \mathcal{W}$, if $b^*_{s,w}(t)=0$, then  $\left(p^*_s(t)),p^*_w(t)\right) = (0,0)$. However, if $b^*_{s,w}(t)=1$, then $\left(p^*_s(t)), p^*_w(t)\right)$ should be the optimal solutions for the power control policy of users $s$ and $w$. In order to facilitate the SIC process at the BS, the strong IoTD is assumed to transmit at the maximum power $P_{\max}$. Meanwhile, the transmit power for weak IoTD needs to be optimized. Therefore, for all $(s,w) \in \mathcal{S} \times \mathcal{W}$, if we assume that the IoTDs are paired together and that we obtain the optimal power allocation $p^*_w(t)$, then problem $\mathcal{OP}$ will simply become a linear assignment that determines the optimal pairing policy $(b^*_{s,w})_{(s,w) \in \mathcal{S} \times \mathcal{W}}$. Here, for all $(s,w) \in \mathcal{S} \times \mathcal{W}$, since we aim to determine the optimal power allocation that satisfies the SINR threshold for each pair of IoTDs such that the AoI is minimized, we can reduce the feasible set of the power control of problem $\mathcal{OP}$ to the set of the power allocation that minimize the sum of AoI of each pair of IoTDs. Consequently, problem $\mathcal{OP}$ can be re-written as:
\begin{subequations} 
\begin{align}
\text{$\mathcal {OP}_{outer} $: } & \min_{ \boldsymbol B} \dfrac{1}{T} \sum_{t<T} \Big[\dfrac{1}{2I} \sum_{s \in \mathcal{S}} \sum_{w \in \mathcal{W}} \Big\{  A_{s}(t) + A_{w}(t) \Big\} b_{s,w}(t) \Big] \\
& \text{s.t.  \,\,\, (\ref{consb1}),
(\ref{consbnf1}), (\ref{consbnf2}), 
}  
\end{align}
\end{subequations}
where $p^*_{w}(t)$ can be obtained by solving the following feasibility check optimization problem
\begin{subequations} 
\begin{align}
\hspace{-0.2cm}\mathcal{O}&\mathcal{P}_{inner}: \mathrm{Find}~~ \boldsymbol{p}_w \\
&\text{s.t.} \,\,\, p_w(t), p_s(t) \leq P_{\max}, \qquad \forall w \in \mathcal{W}, s \in \mathcal{S}, t \in \mathcal{T}, \\
&\gamma_{s}(t) \geq \gamma_{\rm th}, \hspace{2.2cm}\,\,\,\,\,\,  \forall s \in \mathcal{I}, t \in \mathcal{T},\\
&\gamma_{w}(t) \geq \gamma_{\rm th}, \hspace{2.2cm}\,\,\,\,\,  \forall w \in \mathcal{I}, t \in \mathcal{T},
 \end{align}
\end{subequations}
\noindent for each pair of IoTDs $(s,w)  \in \mathcal{S} \times \mathcal{W}$. Consequently, $\mathcal {OP}_{inner} $ is a power allocation problem for a given cluster of IoTDs and it defines the set of feasible solutions for problem $\mathcal {OP}_{outer}$, which is a linear assignment problem. Nonetheless, $\mathcal {OP}_{inner}$ needs to be solved for all possible combinations of IoTDs $\left(s,w\right) \in \mathcal{S} \times \mathcal{W}$. Therefore a computationally efficient approach is presented to solve $\mathcal {OP}_{inner}$ in the following section. 
\subsubsection{Power Control} Let us consider the pair of NOMA IoTDs $(s,w) \in \mathcal{S} \times \mathcal{W}$. The goal here is to obtain a possible value of the power allocation for the weak user $w$, i.e., $p_{w}$, that satisfies the SINR constraints of IoTDs $s$ and $w$ when paired together. 
\begin{subequations}
\begin{align}
&\gamma_{s}(t) \geq \gamma_{\rm th}, \\
&\gamma_{w}(t) \geq \gamma_{\rm th}.
\end{align}
\label{eq:SINR_constraints}
\end{subequations}
From the SINRs expressions in \eqref{snrconstraint} and \eqref{snrconstraint2}, one can conclude that the SINRs constraints in \eqref{eq:SINR_constraints} are satisfied if and only if 
\begin{equation}
    p_{w}^{\min} \leq  p_{w}^{\max},
\end{equation}
where $p_{w}^{\min}$ and $p_{w}^{\max}$ are expressed, respectively, as 
\begin{subequations}
\begin{align}
    p_{w}^{\max} &= \min \left( \frac{ p_s|h^n_{s \rightarrow b}(t) |^2 - \gamma_{\rm th} \sigma^2 }{ \gamma_{\rm th}| \boldsymbol h_{w \rightarrow R}(t) \boldsymbol \Phi(t) \boldsymbol h^H_{R \rightarrow b}(t) + h_{w \rightarrow b}(t) |^2   }, P_{\max} \right), \\
    p_{w}^{\min} &= \frac{\gamma_{\rm th} \sigma^2 }{| \boldsymbol h_{w \rightarrow R}(t) \boldsymbol \Phi(t) \boldsymbol h^H_{R \rightarrow b}(t) + h_{w \rightarrow b}(t) |^2}.
\end{align}
\label{eq:alpha_rng}
\end{subequations}
\noindent
Based on this, any random value of the power $p_{w}$ within the range $[p_{w}^{\min},p_{w}^{\max}]$ is a feasible value for problem $\mathcal{OP}_{inner}$. 
\subsubsection{User clustering} After obtaining the optimal power allocation for each possible NOMA cluster along with its corresponding sum AoI, we can apply the \textit{Hungarian method} to determine the optimal clustering configuration \cite{9140040}. The main motivation behind this is that the \textit{Hungarian method} can optimally solve the 2-dimensional matching problem. The input of the Hungarian method is an $I\times I$ cost matrix, which is the sum of AoI of the strong and weak IoTDs in each cluster, and its output is the pairing matrix $\boldsymbol B^*$ where $\boldsymbol B^*(s,w)$ = 1 shows that strong user $s \in \mathcal{S}$ is paired with weak user $w \in  \mathcal{W}$, and $\boldsymbol B^*(s,w)$ = 0 otherwise.  
\subsubsection{Complexity Analysis} 
The phase-shift matrix problem is a semi-definite programming (SDP) problem that can be solved by the interior point method and its order of computational complexity with $m$ SDP constraints that contain an $n \times n$ positive semidefinite matrix is given as $\mathcal{O}(\sqrt{n}\text{log}(1/\epsilon)(mn^3 + m^2n^2 + m^3))$, where $\epsilon>$0 is the solution accuracy \cite{Elhattab_JSAC}. For $\mathcal{OP}$1, with $m$ = $L$ and $n$ = $L+1$, the approximate computational complexity to solve SDP can be written as $\mathcal{O }(\text{log}(1/\epsilon)(L^{4.5})$. Meanwhile, let $x$ be the maximal number of generated Gaussian random vectors and $T_{GR}$ is the complexity of performing one Gaussian random iteration. The approximate complexity of $\mathcal{OP}$1 can be written as $\mathcal{O }(\text{log}(1/\epsilon)(L^{4.5}+ xT_{GR}))$. Meanwhile, the computational complexity of calculating total transmit power is approximately $\mathcal{O}(1)$. The power computations complexity for $I^2$ user combinations is $\mathcal{O}(I^2)$. We then employ the Hungarian algorithm, which has polynomial time complexity of $\mathcal{O}(I^2)$ for the assignment problem. As a result, the approximated total complexity of our solution is $\mathcal{O }(\text{log}(1/\epsilon)(L^{4.5} + xT_{GR}))$.

\begin{figure*}[ht]
\centering
\subfigure[Impact of number of RIS elements.] {\centering\includegraphics[width=0.24\textwidth]{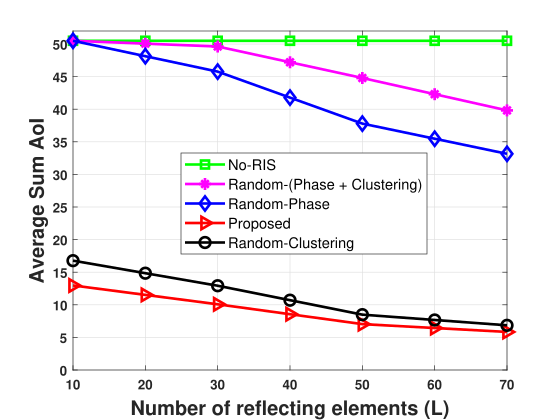}} 
\subfigure[Average age per IoT device.]
{\centering\includegraphics[width=0.24\textwidth]{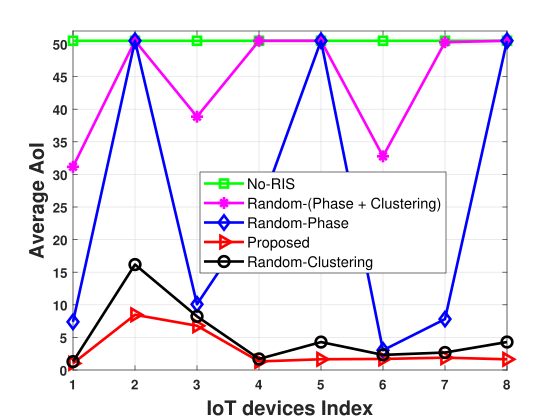}}
\subfigure[Impact of SINR Threshold on AoI.]
{\centering\includegraphics[width=0.24\textwidth]{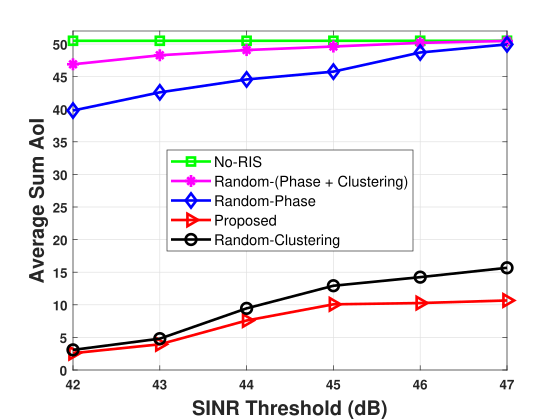}}
\subfigure[Impact of Power on AoI.] {\centering\includegraphics[width=0.24\textwidth]{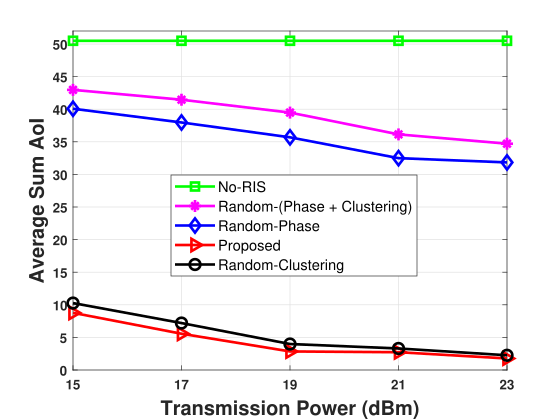}} 
\caption{Optimizing the user clustering, power allocation and phases of RIS elements}
\label{results}
\end{figure*}

 \section{Simulation and Numerical Analysis}
  \label{simulationresults}
We consider a 3-D area that consists of one BS, one RIS and a set of IoTDs. We assume that the global coordinate system $(X, Y, Z)$ is Cartesian. The BS is located at $(0,0,H_{\rm b})$ and the RIS is located at $(d_{R \rightarrow b},0,H_{\rm r})$, where $d_{R \rightarrow b}$ is the distance from RIS to the BS, and $H_{\rm b}$ and $H_{\rm r}$ are the heights of the transmit antenna of the BS and of the RIS, respectively. In addition, multiple IoTDs are randomly distributed at the ground level, where for all $i \in \mathcal{I}$, the locations of IoTDs are $(x_i,y_i,0)$. We have two sets of IoTDs, i.e., the set of strong IoTDs $\mathcal{S}$ and the set of weak IoTDs $\mathcal{W}$. The strong IoTDs are randomly located within a circular area centered at the BS and with radius $d^s$. Thus, the coordinates of the strong IoTDs are given by $x^s_i = d^s{\cos(\theta^s_i)}$ and $y^s_i = d^s\sin(\theta^s_i)$, where $\theta^s_i \in [0, 2 \pi]$ is a polar angle. On the other hand, the weak IoTDs are randomly located within a circular area centered at the RIS and with radius $d^w$. Thus, coordinates of the weak IoTDs are given by $x^w_i = d^w_{i}\cos(\theta^w_i)$ + $d_{R\rightarrow b}$ and $y^w_i = d^w_{ i}\sin(\theta^s_i)$, where $\theta^w_i \in [0, 2 \pi]$ is a polar angle. The total number of time-slots $T=100$ and number of IoTDs $I=20$, while the communication parameters are taken as $d^s_i=10$m, $d^w_{i}=10$m, $d_{R \rightarrow b} =150$m, $\sigma^2-110$ dBm,  $\eta_{R \rightarrow b}-2.2$, $\eta_{w \rightarrow R}=-2.2$, $\eta_{s \rightarrow b}$= $\eta_{w \rightarrow b}$ = -3.5, $K1$ = 2, $H_{\rm b}$=  $H_{\rm r}=10$m,  $\gamma_{\mathrm{th}}=45$dB. All the results are averaged over $500$ Monte Carlo experiments. 
\par In order to evaluate the performance of the proposed algorithm, we compare its performance with the following three baselines schemes: 
\begin{itemize}
\item Optimal clustering with random RIS configuration: In this scheme,
a random phase-shift is determined for the weak IoTDs, whereas the user clustering and the power allocation are performed using the proposed approach.
\item Random clustering with optimized RIS configuration: In this scheme, a random clustering is formed between the strong and the weak IoTDs, whereas the RIS phase-shift matrix and the power allocation are optimized using the proposed approach.
\item Random clustering with random RIS configuration: In this scheme, the users clustering and the RIS phase-shift matrix are both random.
\end{itemize}
\par Fig.~\ref{results}(a) depicts the impact of the size of the RIS on the average sum AoI. It can be seen that the RIS has a significant impact on the AoI and increasing the number of the RIS elements results in decreasing the AoI for all schemes. This is evident since the RIS helps to improve the channel quality of the weak IoTDs which leads to augmenting the likelihood of pairing with strong IoTDs and eventually ends up decreasing the AoI. In addition, we observe that the proposed algorithm achieves the lowest sum AoI compared to the other approaches. Particularly, the proposed scheme achieves around $88.43\%$, $85.33\%$, and $82.39\%$ decrease in the average AoI with $70$ RIS elements compared to the case when the RIS is not deployed, when the RIS configuration is random and the user clustering is optimal, and when both the RIS configuration and the user clustering are random, respectively. It can be observed that as the number of RIS elements are increased, the gap in the AoI values obtained with the random and optimal user clustering schemes shrink, e.g., $22.69\%$ with $L=10$, and $14.79\%$ with  $L=70$. The reason behind is that with a large number of RIS elements, the channel gains of the weak IoTDs increase and, therefore, the random pairing also achieves lower AoI values. Another observation that can be made regarding the importance of the optimal clustering when relying on random RIS configuration. Considering a random RIS phase shift matrix, optimizing the user clustering will be indispensable. e.g, the AoI achieved by the random clustering gets around $16.72\%$ higher than the one with the optimal clustering with $L = 70$ RIS elements.
\par We plot the average AoI for a set of IoTDs in Fig.~\ref{results}(b). It can be seen that the proposed method has a lower average sum AoI per user compared to the other approaches. Moreover, the average AoI gap is relatively high among different schemes, which highlights the significance of the optimal RIS configuration, user clustering and power allocation. In addition, it is observed that the random phase method decreases the AoI for some devices and increases it for other devices. 
\par Fig.~\ref{results}(c) depicts the impact of the SINR threshold on the average sum AoI. It can be seen that, as the SINR threshold increases, the average sum AoI starts increasing for all the schemes. However, the proposed method achieves the lowest sum AoI compared to the others. Another observation can be seen when the SINR threshold is low. In such a case, the proposed approach and the random user clustering give close AoI values but as the SINR threshold increases, the gap start increasing and becomes almost double (from $16.39\%$ to $31.96\%$) when the SINR threshold is increased from $42$ to $47$ dB. The approaches of the random RIS configuration with optimal and random clustering provide the worst AoI values which reinforces that the  optimization of both the RIS configuration and the user clustering are important to achieve the required freshness of information.  Finally, Fig.~\ref{results}(d) shows the impact of the maximum transmit power per user on the average sum AoI. It can be observed that when the user transmit power increases, the SINR increases and, hence, the AoI decreases. 
\section{Conclusion}
We explored the integration of RIS in NOMA-based IoT networks to preserve information freshness. An optimization problem is formulated to minimize the average sum AoI by optimizing the user-clustering, user transmit power and the RIS configuration. To tackle this, RIS configuration problem is solved first using the SDR approach. Then, the joint user-clustering and power control problem is decomposed into disjoint user-clustering and power-control. The Hungarian method is adapted to solve the user-clustering sub-problem while the feasible transmit power range is obtained for the power-control sub-problem. Numerical results demonstrate that the proposed method has superior performance.
\label{conclusion}
 \bibliographystyle{IEEEtran}
 \bibliography{IEEEabrv,biblio}
\end{document}